# Comment on "Route from discreteness to the continuum for the Tsallis $q$-entropy"

Congjie Ou[1] and Sumiyoshi Abe[1,2,3]

[1]*Physics Division, College of Information Science and Engineering, Huaqiao University, Xiamen 361021, China*
[2]*Department of Physical Engineering, Mie University, Mie 514-8507, Japan*
[3]*Institute of Physics, Kazan Federal University, Kazan 420008, Russia*

Several years ago, it has been discussed that non-logarithmic entropies such as the Tsallis $q$-entropy cannot be applied to systems with continuous variables. Now, in their recent paper [Phys. Rev. E **97**, 012104 (2018)], Oikonomou and Bagci have modified the form of the $q$-entropy for discrete variables in such a way that its continuum limit exists. Here, it is shown that this modification violates the expandability property of entropy, and their work is actually a supporting evidence for the absence of the $q$-entropy for systems with continuous variables.

As often pointed out, "entropy" has become a confusing terminology since it has been used in information theory. A lot of efforts have been devoted to improve the situation, and today there is a consensus that "information" is a physical quantity (see Refs. [1-3], for example). However, the concepts still seem to remain ambiguous if generalized entropies are considered. Accordingly, the axiomatic approach to information theory may sometimes be useful, and in fact it constitutes the main point of the present comment.

Statistical mechanics tells us that the Clausius entropy, $S$, in thermodynamics is related to the number of microspocically accessible states, $W$, as $S = k_B \ln W$, which is the celebrated Boltzmann relation that connects the macroscopic world with the microscopic one, where $k_B$ is the Boltzmann constant and henceforth is set equal to unity for the sake of simplicity. This relation can be obtained from the quantity,

$$S \equiv S^{(W)} = -\sum_{i=1}^{W} p_i \ln p_i ,  \qquad (1)$$

in a special case when the probabilities of finding the system in all states are identical: $p_i = 1/W$ ($i = 1, 2, ..., W$). $S$ in Eq. (1) is formally equivalent to the Shannon entropy in information theory.

From these, it is clear that the entropy is primarily concerned with countable sets. In physics, however, many important systems are formulated in continuous (phase) space. Therefore, to define the entropy there, it is essential to perform coarse graining. The scale of coarse graining in the case of phase space ($\Gamma$ space) is given by the Planck constant. Thus, the Planck constant is indispensable even in classical statistical mechanics [4]. Shift from a discrete set to a continuum one is, however, not so straightforward. As noted in Ref. [5], such a procedure requires a sensitive discussion of measure theory.

In information theory, the shift from discrete to continuum is rather formal. The resulting quantity is written as $S = -\int dx\, p(x) \ln p(x)$, which is called the differential entropy [6], where $p(x)$ is the probability density, that is, $p(x)dx$ is the probability that the random variable, $X$, of the system is realized in an infinitesimal interval, $[x, x+dx]$. (Here, we are considering a case of a single variable for simplicity.) The differential entropy is defined in the continuum limit of the one analogous to Eq. (1) with a uniform measure on the original discrete set with the scale of coarse graining being set equal to unity. However, the situation becomes involved if a generalization of Eq. (1) is considered.

Several years ago, it has been shown [7] (see also Ref. [8]) that generalized entropies of the non-logarithmic form do not have the continuum limits, in general. This result has put a stringent limitation on the use of such entropies in statistical mechanics. However, now the authors of Ref. [9] discuss that if the non-logarithmic Tsallis $q$-entropy [10] indexed by $q$ ( $>0$ ), $S_q^{(n)} = \sum_{i=1}^{n} p_i \ln_q(1/p_i)$ with $\ln_q x = (x^{1-q} - 1)/(1-q)$ ($x > 0$), is modified as

$$\tilde{S}_q^{(n)} = n^{q-1} \sum_{i=1}^{n} p_i \ln_q(1/p_i), \tag{2}$$

then the continuum limit $n \to \infty$ exists, and the relative entropy of the Csiszár-type, which is a generalization of the Kullback-Leibler divergence, is obtained, provided that $W$ in Eq. (1) is rewritten here as $n$ in order to adjust the notation to that used in Ref. [9].

Now, our comment on Ref. [9] is the following. As widely accepted, entropy should satisfy the *expandability* property stated as follows:

$$S^{(n+1)}(p_1, p_2, ..., p_n, p_{n+1}=0) = S^{(n)}(p_1, p_2, ..., p_n), \qquad (3)$$

which is employed as an axiom for characterizing the Shannon entropy [11] and the $q$-entropy [12]. This property is natural: entropy of a system having $n$ states is identical to that of a system having $n+1$ states with the last state being never occupied.

Clearly, the quantity in Eq. (2) violates this property due the prefactor, $n^{q-1}$. In addition, if this point is considered in the context of statistical mechanics, then it necessarily implies that the Boltzmann constant (being set equal to unity, here) has to depend on the number of states, leading to a physically unacceptable result. In other words, the authors of Ref. [9] actually present a new supporting evidence for the result presented in Ref. [7]: the $q$-entropy cannot be defined for continuous variables.

In conclusion, we have examined Ref. [9] in view of the expandability axiom for entropy and have shown that it further illustrates the absence of the continuum limit of the $q$-entropy.

The work of CO was supported by the grants from Fujian Province (No. 2015J01016, No. JA12001, No. 2014FJ-NCET-ZR04) and from Huaqiao University (No. ZQN-PY114). SA was supported in part by a grant from National Natural Science Foundation of China (No. 11775084) and Grant-in-Aid for Scientific Research from the Japan Society for the Promotion of Science (No. 26400391 and No. 16K05484), and by the Program of Competitive Growth of Kazan Federal University from the Ministry of Education and Science of the Russian Federation.

——————————